\documentclass[%
 pra, 
 superscriptaddress,
 showpacs,
 longbibliography,
 aps,
 showkeys]{revtex4-2}

\usepackage[pdftex]{hyperref,color,graphicx}
\usepackage{amsfonts,amssymb,amsmath}
\usepackage[separate-uncertainty=false]{siunitx}
\usepackage[english]{babel}
\usepackage[utf8]{inputenc}
\usepackage[T1]{fontenc}
\usepackage[toc,page]{appendix}
\usepackage{graphicx}

\usepackage[space]{grffile}
\usepackage{latexsym}
\usepackage{textcomp}
\usepackage{longtable}
\usepackage{multirow,booktabs}
\usepackage{url}
\hypersetup{colorlinks=false,pdfborder={0 0 0}}
\usepackage[english]{babel}
\usepackage{dcolumn}
\usepackage{bm}
\usepackage{units}
\usepackage{upgreek}
\usepackage{color}
\usepackage{cancel}
\usepackage{hyperref}
\usepackage{braket}
\usepackage{epstopdf}

\newcommand{\bbZ}{\mathbb{Z}}

\begin{document}

\title{Parrondo's paradox for discrete-time quantum walks in momentum space}

\author{Georg Trautmann}
\affiliation{Institut f\"u{r} Theoretische Physik, Universit\"{a}t Heidelberg, Philosophenweg 16, 69120 Heidelberg, Germany}

\author{Caspar Groiseau}
\affiliation{Departamento de F\'isica Te\'orica de la Materia Condensada and Condensed Matter Physics Center (IFIMAC), Universidad Aut\'onoma de Madrid, 28049 Madrid, Spain}

\author{Sandro Wimberger}
\email{sandromarcel.wimberger@unipr.it}
\affiliation{Dipartimento di Scienze Matematiche, Fisiche e Informatiche, Universit\`{a} di Parma, Parco Area delle Scienze 7/A, 43124 Parma, Italy}
\affiliation{INFN, Sezione di Milano Bicocca, Gruppo Collegato di Parma, Parco Area delle Scienze 7/A, 43124 Parma, Italy}

\begin{abstract}
We investigate the possibility of implementing a sequence of quantum walks whose probability distributions give an overall positive winning probability, while it is negative for the single walks (Parrondo's paradox). In particular, we have in mind an experimental realisation with a Bose-Einstein condensate in which the walker's space is momentum space. Experimental problems in the precise implementation of the coin operations for our discrete-time quantum walks are analysed in detail. We study time-dependent phase fluctuations of the coins as well as perturbations arising from the finite momentum width of the condensate. We confirm the visibility of Parrondo's paradox for experimentally available time scales of up to a few hundred steps of the walk.
\end{abstract}

\keywords{quantum walk, Parrondo's Game, game theory, Hamiltonian ratchets, Bose-Einstein condensates, noisy quantum systems, atom-optics kicked rotor}

\maketitle

\section{Introduction}
\label{Intro}

Parrondo's paradox describes a situation in which a winning strategy is obtained even if the comprehensive game consists of various combinations of single losing games. In other words, each game for itself is losing on average whilst an appropriate combination of them may be winning on average \cite{Parrondo1, Parrondo2, Harmer1, Harmer2, CQP-review}. 
The apparent paradox was shown to be important in many biological and physical systems, but can also be observed when playing a classical game with coins \cite{CP0, CP1}.
Classical coin games have their analogue in quantum mechanics, in so-called discrete-time quantum walks (DTQW) \cite{Portugal}, first introduced in 1993 \cite{QRW}. Quantum versions of Parrondo's paradox have indeed been proposed, see e.g. \cite{QP1, QP2, QP3, CQP-review, Game, Ratchet}, and we are aware of at least one experimental realisation \cite{AdvQT}.

In the classical version of Parrondo's paradox it typically originates in built-in correlations between the two (or more) walks of the sequence \cite{Parrondo1, Parrondo2, Harmer1, Harmer2}. Due to coherent interference and the entanglement of the states in DTQWs quantum versions of Parrondo's paradox support, in principle, much richer behavior \cite{LMP2016, QP0, FNL}, and even uncorrelated walk sequences show the effect, at least for some finite evolution time \cite{QP1, QP2, QP3, Game, Ratchet, AdvQT}. We are not aware of a general statement on the existence of necessary and sufficient conditions for the occurrence of the paradox in the quantum case, but refer to  \cite{LMP2016, CQP-review} for comprehensive overviews.

In this paper, we are revisiting the experiment reported in \cite{AdvQT} which was performed with single photons. We check whether a similar experiment could, in principle, be performed using a Bose-Einstein condensate (BEC) which is subject to kicks in momentum space. The latter realisation was successfully used so far to implement DTQWs up to about 20 steps  \cite{PRL2018, PRA2019, PRR2021}. The coin operation in this experiment is done shining microwave (MW) pulses onto the atoms of the condensate of which two internal states act as a qubit or spin-1/2 coin-state space. Such MW pulses have to be precisely timed and controlled which is often not so simple. We investigate here how the proposal of \cite{AdvQT} can be simplified and better adapted to the BEC setup and whether it turns out to be robust with respect to experimental uncertainties in the phase of the MW. In addition, the finite width of the BEC in (quasi)momentum space is considered. We will show that one may indeed implement Parrondo's paradox with the BEC setup for a substantial number of steps. Having at hand such a robust experimental platform for the realization of
a quantum Parrondo's paradox could open the road for investigating novel protocols for, e.g., entanglement creation, quantum cryptography or quantum chaotic random number generators \cite{LC2021, WB2021}.

\section{The experimental setup}
\label{model}

We are looking at a one-dimensional DTQW \cite{Portugal} in the following, as it was realised in \cite{PRL2018, PRA2019, PRR2021} using a rubidium Bose-Einstein condensate.
The Hilbert space for this system is given by $\mathcal{H} = \mathcal{H}_C \otimes \mathcal{H}_P $, where $\mathcal{H}_C$ is the two-state coin space, spanned by the orthonormal basis $\{\ket{0}_C,\ket{1}_C\}$(corresponding to the two  ground state hyperfine levels of rubidium) and $\mathcal{H}_P$ represents the infinite walker's space spanned by $\{\ket{n}; n \in \mathbb{Z} \}$ (corresponding to the center-of-mass momentum of the rubidium Bose-Einstein condensate).
Each step of the walk consists of two operations, the rotation in the coin space $ \mathcal{H}_C$ by a coin operator $M$ and a kick onto the walker by an operator $\hat U$. The latter describes the movement of the walker in momentum space based on the internal state, which produces a strong entanglement between the walker's degrees of freedom in each step. After iteratively applying a certain number of steps $t$, the state of the Bose-Einstein condensate can be described as \(\ket{\Psi (t)} = (\hat{U}M)^t \ \ket{\Psi (0)}\), where \(\ket{\Psi (0)}\) is the initial state.
The general coin operator $M$ is the one used in \cite{AdvQT}, 
\begin{equation}
\label{eq:1}
    M(\alpha,\gamma, \chi) = 	
    \begin{pmatrix}
    e^{i \alpha}\cos\chi & \ -e^{-i \gamma}\sin\chi \\
    e^{i \gamma}\sin\chi & \ e^{-i \alpha}\cos\chi 
    \end{pmatrix},
\end{equation}
with three real parameters $\alpha, \gamma$ and $\chi$ (the "Euler" angles of the spin-state rotation) and is realized with microwave radiation addressing the hyperfine levels of rubidium.
In contrast to \cite{AdvQT} we do not study an ideal quantum walk with exclusively nearest-neighbor couplings but a realisation based on the quantum kicked rotor
\cite{AdvAtMol, PhysRep90, PhysRep22} with a Bose-Einstein condensate subject to time-periodic kicks from a standing-wave laser \cite{PRL2018, PRA2019, PRR2021}.

The dynamics of the one-dimensional kicked rotor model with assumed $\delta$-like kicks is induced by the Hamilton operator \cite{AdvAtMol}
\begin{equation}
\label{eq:2}
\hat H(t) = \frac{\hat p^2}{2} + k  \cos(\hat{\theta}) \sum_{j\in \bbZ} \delta (t-j \tau) ,
\end{equation}
with the position (angle) and momentum operators $\hat \theta$ and $\hat p$, the kicking strength $k$, the kicking period $\tau$, and  the kick number $j$.
Using Bloch's theorem the line operator $\hat x$ of the experiment working with periodic optical lattices can be reduced to an angle $\hat \theta$ with periodic boundary conditions \cite{AdvAtMol}. The corresponding evolution of the quantum system, $i\frac{\partial}{\partial t} \ket{\Psi(t)} = \hat{H} (t) \ket{\Psi(t)}$, can be described by the \textit{Floquet operator} $\hat{U} (t)$ \cite{AdvAtMol}.
Hence, in a quantum walk based on the kicked-rotor model, the shift operation is replaced by a kick impacted by the Floquet operator,
\begin{equation}
\label{eq:3}
    \hat U =  \hat K \hat F \equiv e^{-i k \sigma_z \cos(\hat{\theta}) } e^{-\frac{i}{2}\tau(\hat{n}+\beta)^2} .
\end{equation}
One step of the walk is then given by the subsequent application of the coin operator $M$ and $\hat U$. 
The first factor of the Floquet operator $\hat U$ describes the kick on the atoms by $\hat K = e^{-i k  \sigma_z \cos(\hat{\theta})}$ 
and the second factor is the free time evolution between two kicks by $\hat F = e^{-\frac{i}{2}\tau \hat p^2}$. Due to Bloch's theorem the momentum can be decomposed into 
$\hat p = \hat n + \beta$ with the integer part $\hat{n}$ and the quasimomentum $\beta \in [0;1)$ in our dimensionless variables.
If $\beta = 0$ and $\tau=4\pi$, we see immediately that the Floquet operator only consists of the kicks without any phase evolution. This is the so-called
quantum resonant regime of the kicked rotor \cite{AdvAtMol, PhysRep90, PhysRep22}. 
In this regime the evolution well approximates an ideal quantum walk as suggested in 
\cite{PRA2016a} and subsequently realised in \cite{PRL2018, PRA2019, PRR2021}, in particular for kick strengths of the order $k \approx 1.5$, for which the
kicks dominantly couple nearest-neighbor momentum states. There is another important difference to ideal quantum walks: the kicked-rotor dynamics for both internal states would be completely symmetric in momentum space if the parity symmetry was not broken. This symmetry is best broken by starting from an asymmetric initial state in momentum space, realising ratchet dynamics, and discussed in great detail in \cite{PRA2016, AP2017, Mark2007}. The atoms are hence prepared in a "ratchet" state in momentum space of the following form
\begin{equation}
    \label{eq:4}
    |\psi_k (0)\rangle = \frac{1}{\sqrt{S}} \sum_{s} e^{-i s \pi/2} \ket{n=s},
\end{equation}
with $s = \ldots -2, -1, 0, 1, 2, \ldots$ and $S$ being a normalization factor that equals the number of contributing states. These states move now in a preferred direction depending on the sign of the kick, see the Pauli matrix $\sigma_z$ in Eq. \eqref{eq:3}. This is achieved by tuning the laser kicking the atoms between the two coin-states, so that they experience detunings of opposite sign, which translates into said sign difference in the kick strength. Hence, the two internal states move into different directions \cite{PRA2016a, PRA2016, AP2017}. The coin operator, Eq. \eqref{eq:1}, mixes those states depending on the choice of the angle parameters. The quantum walk is hence governed essentially by those angles. Please note that a Parrondo's game based on a similar kicked-rotor ratchet was proposed in \cite{Ratchet}, lacking however, the coin degree of freedom of our DTQW realization.

\begin{figure}[th]
\centering
\includegraphics[width=0.75\textwidth]{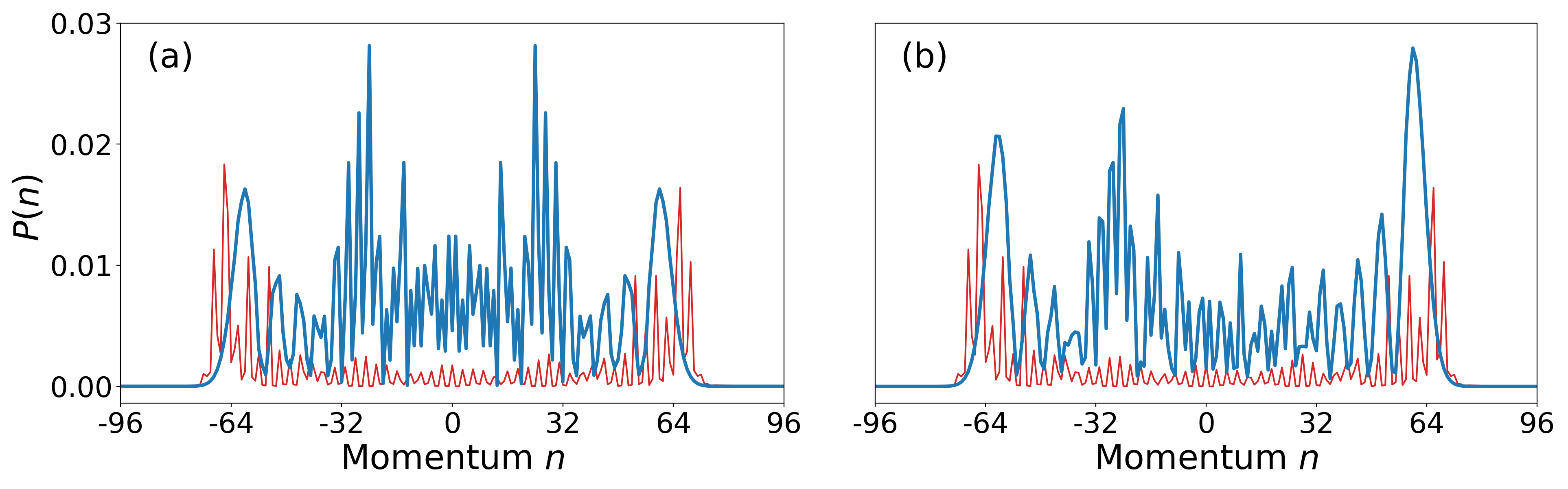}
\caption{Momentum distributions $P(n)$ of walk A (after $N=50$ steps) for $(a)$ an initial state with only one momentum state at $n = 0$ and $(b)$ a ratchet state using three momentum states. The kicked-rotor walks are displayed by the thick blue distributions whereas the thin red lines display ideal quantum walks with only nearest-neighbor couplings. The perfect parity symmetry in the blue distribution of $(a)$ is the result of not preparing the atom in a ratchet state and is therefore not further useful for us. Although the blue part  in $(b)$ shows differences with respect to the ideal walk, it still shows qualitatively a similar behaviour what concerns the (here negative) winning probability $O(N)$.}
\label{fig:1}
\end{figure}

For our numerical simulations, we mostly used the initial state
\begin{equation}
\label{eq:5}
 \ket{\Psi(0)} = \frac{1}{\sqrt{6}}(\ket{0}+i\ket{1})_C \otimes (-i \ket{n = -1} + \ket{n = 0} +i \ket{n = 1})_P,
\end{equation}
whose evolution turned out to be closest to an ideal quantum walk. Figure \ref{fig:1} shows a direct comparison for a walk with a coin matrix 
$M_A(137.2^\circ, 29.4^\circ, 52.1^\circ)$, from now on denoted walk A.

\section{Previous Proposal}
\label{china}

\begin{figure}[th]
\centering
\includegraphics[width=0.6\textwidth]{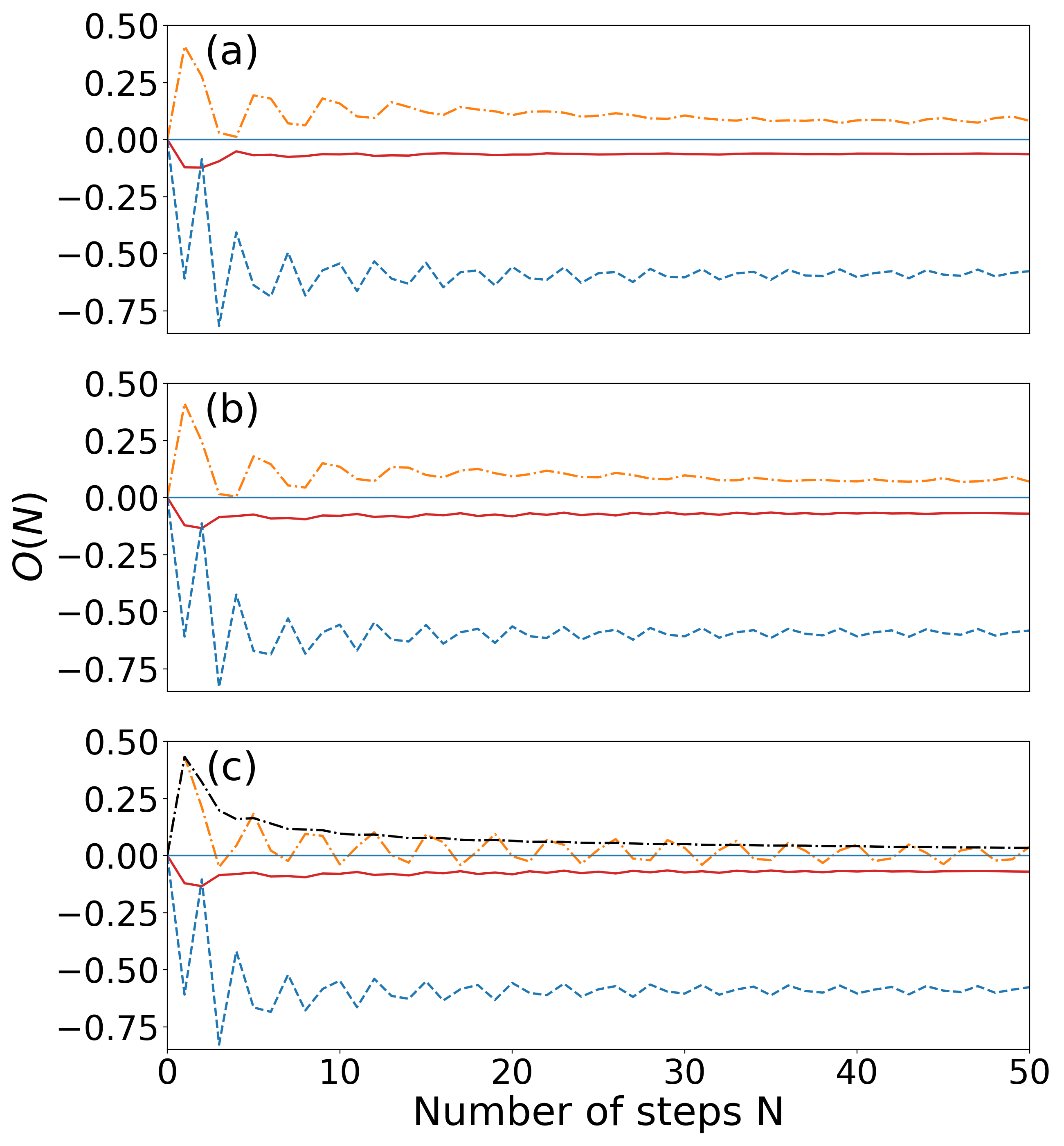}
\caption{Plots of the winning probability $O(N)$ for every walk vs. step number $N$ for $k=1.56$. In all three panels the winning probability of walk A is represented by the solid red line, of walk B by the dashed blue line, and of the walk ABB by the dash-dotted orange line. Only in $(c)$ there is an additional observable, the time-averaged winning probability of the walk ABB $\overline{O(N)}$ (black dash-dotted). (a) displays the situation with $M_A(137.2^\circ, 29.4^\circ, 52.1^\circ)$ and $M_B(149.6^\circ, 67.4^\circ, 132.5^\circ)$, whereas (b) shows the outcomes with only $\alpha_A' = 0$ but $\alpha_B' = \alpha_B - \alpha_A = 12.4$, i.e. $M_A(0, 29.4^\circ, 189.3^\circ)$ and $M_B(12.4^\circ, 67.4^\circ, 269.7^\circ)$. In (c) both coins A and B have its $\alpha_{A,B}=0$ implying $M_A(0, 29.4^\circ, 189.3^\circ)$ and $M_B(0, 67.4^\circ, 282.1^\circ)$.}
\label{fig:2}
\end{figure}

Analogous to \cite{AdvQT} we study quantum walks consisting of two different walks, a walk A and a walk B, which form the periodic sequence ABB. All our walks are carried out with kicking strength $k=1.56$ and the initial state of Eq. \eqref{eq:5} for typically up to $N = 50$ steps. This is sufficient since most experimental implementations of quantum walks go up to a few tens of steps, see e.g. \cite{AdvQT, PRL2018, PRA2019, Walk-Selection}. Following \cite{AdvQT}, we observe the winning probability after $N$ steps
\begin{eqnarray}
\label{eq:obs}
O(N) \equiv P_R(N) - P_L(N) \ \ \ \textrm{with} \ \ \ P_L = \sum_{n = - \infty}^{-1}\braket{\Psi(N)}{(I_C\otimes|n\rangle\langle n|)|\Psi(N)} \\ 
\textrm{and} \ \ \ P_R = \sum_{n =  1}^{\infty}\braket{\Psi(N)}{(I_C\otimes|n\rangle\langle n|)|\Psi(N)}.
\end{eqnarray}
A walk or game is considered as a winning game if this probability is positive $O(N) > 0$. The winning probability is just the integrated momentum distribution, weighting the positive/negative walker's positions as positive or negative, respectively. Parrondo's paradox is satisfied when the winning probability of a specific sequence of walks A and B has an opposite sign with respect to the two individual walks \cite{Harmer1, Harmer2}.  The coin operators used in \cite{AdvQT} are $M_A(137.2^\circ, 29.4^\circ, 52.1^\circ)$ and $M_B(149.6^\circ, 67.4^\circ, 132.5^\circ)$, making an ideal quantum walk A as well as B losing with negative $O(N)$. The sequence ABB is, however, winning with positive $O(N)$. Note, that it is not required for walk B to produce a larger value of $O(N)$ compared to walk A (see Fig. \ref{fig:2}) in order to observe the paradox. In \cite{rsos}, a quantum version of Parrando's paradox was proposed with a higher absolute value of $O(N)$ in walk A, however, with a different sequence of single walks. According to a theorem reported in  \cite{FNL, ChinPhysB}, the outcome of the walks and also of the game ABB does only depend on the sum of the two angles $\alpha$ and $\gamma$, while their individual values are not important. This is only true, however, for the specific initial state $\ket{\Psi(0)} = \frac{1}{\sqrt{2}}(\ket{0}+i\ket{1})_C \otimes (\ket{n = 0})_P$, in our language, and for ideal walks. 

\section{Adaption to BEC experiment}
\label{BEC}

In the experiments reported in \cite{PRL2018, PRA2019, PRR2021}, only the angle $\beta$ was well controlled in the MW pulses, usually chosen $\chi = \pi/4$ for unbiased quantum walks. The phase $\gamma$ typically was fixed by a try-and-error procedure in order to ensure symmetrically evolving, i.e. unbiased walks, whilst the  phase $\alpha$ was not explicitly considered. Hence, it would be an important simplification if we may use the theorem mentioned in the previous section in order to eliminate one of the two phase angles.  As discussed before, it is essential for the BEC realization to prepare the atoms in a ratchet state $\ket{\Psi(0)} = \frac{1}{\sqrt{6}}(\ket{0}+i\ket{1})_C \otimes (-i \ket{n = -1} + \ket{n = 0} +i \ket{n = 1})_P$. Moreover, the BEC walks are not ideal, and hence the mentioned theorem does not necessarily apply. Nevertheless, we simulated sequences with the coin-parameter $\alpha$ set to zero and $\gamma$ adjusted accordingly, i.e. using effectively $\gamma ' = \gamma + \alpha$. This simplifies the experimental implementation that would only need to control two of the three angles. Figure \ref{fig:2} shows a comparison of the single A and B walks, as well as the ABB walks for different combinations of $\alpha$ and $\gamma$. When both $\alpha_A=0=\alpha_B$, the winning probability $O(N)$ for each individual step is not necessarily positive any more. Figure \ref{fig:2}(c) shows how it oscillates hitting small negative values. However the integrated, time-averaged winning probability
 \begin{equation}
 \label{eq:6}
    \overline{O(N)} = \overline{P_R(N) - P_L(N)} = \frac{1}{N} \sum_{j=0}^{N} \left(P_R(N) - P_L(N) \right)
\end{equation}
remains positive during the entire walk, see Fig. \ref{fig:2}(c). Actually, it turns out that this time-averaged probability is a much more robust observable also with respect to small fluctuations, which suggests that we should use it in future experiments. We attribute the fact that the theorem mentioned above does not remain exactly valid to the non-ideal kicked-rotor evolution with more than nearest-neighbor couplings and also a finite probability to remain at the initial position (the lazy part of the walk), see \cite{PRA2019, PRA2015} for details.

The proposed original coin rotations, given by $M_A(137.2^\circ, 29.4^\circ, 52.1^\circ)$ and $M_B(149.6^\circ, 67.4^\circ, 132.5^\circ)$, are seemingly optimal for the ideal quantum walks investigated in \cite{AdvQT}. In the case of the kicked-rotor walks, by testing against other coin operators, we found that they produce perfect outcomes as well. However, since the theorem mentioned in Sect. \ref{china} is only approximately valid for our setup, we cannot yet conclude that $M_A(0, 29.4^\circ, 189.3^\circ)$ and $M_B(0, 67.4^\circ, 282.1^\circ)$  give the best possible outcomes. By a brute-force numerical optimization of the coin-operators we were indeed able to find the new matrix $M_A (0, 184.32^\circ, 246,96^\circ)$, for which the integrated winning probability $\overline{O(N)}$ in both walks A and ABB is higher in absolute value compared to the respective walks using the old matrix $M_A(0, 29.4^\circ, 189.3^\circ)$. In walk B, on the other hand, $M_B (0, 67.4^\circ, 282.1^\circ)$ already turned out to produce optimized outcomes. As these optimized coin operators are more likely to be stable under noise, we use $M_A (0, 184.32^\circ, 246,96^\circ)$ and $M_B (0, 67.4^\circ, 282.1^\circ)$ in the following.

In the kicked-rotor walks realised in \cite{PRL2018, PRA2019, PRR2021} their is actually an additional problem arising from a constant energy shift in the Hamiltonian that is different for the two internal spin states of the atoms. This so-called light shift effect induces an additional relative phase of $2k$ between the two states \cite{PRA2019b} any time a kick $\hat K$ occurs, see Eq. \eqref{eq:3}. One may exploit $\alpha$ in our coin matrix now in order to compensate exactly this phase, choosing it $\alpha=k=1.56$ (rad units) in our case \cite{PRA2019b}. Also in this respect it would be of advantage to have $\alpha$ still free, i.e. effectively not used in the single walks A and B as discussed above ($\alpha=0$) such that is can indeed be used as an independent parameter for the compensation of the light shift.

\begin{figure}[th]
\centering
\includegraphics[width=0.6\textwidth]{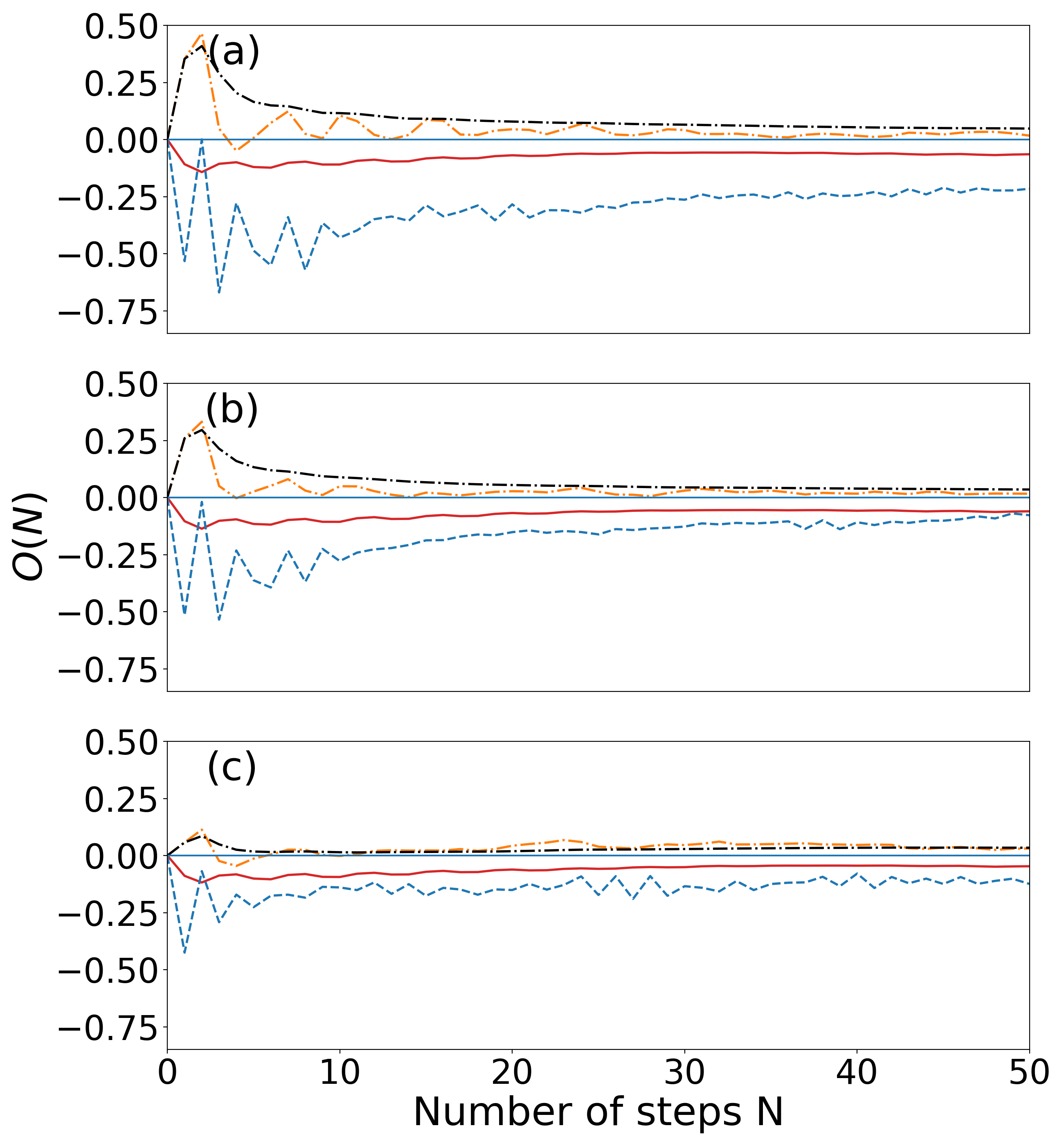}
\caption{$O(N)$ of walk A (solid red lines), walk B (dashed blue lines), and walk ABB (dash-dotted orange lines) as well as the time-averaged winning probability $\overline{O}(N)$ of the walk ABB (dash-dotted black line), each averaged over $R=50$ realisations of the noise. The panels show three different noise-strengths on the coin parameter $\gamma$ with $M_A(0, 184.32^\circ, 246.96^\circ)$ and $M_B(0, 67.4^\circ, 282.1^\circ)$: (a) $\delta(N) \in [- \frac{\pi}{10}, + \frac{\pi}{10}]$, (b) $\delta(N) \in [- \frac{\pi}{5}, + \frac{\pi}{5}]$, and (c) $\delta(N) \in [- \frac{\pi}{3}, + \frac{\pi}{3}]$. The paradoxical behavior of the walks is clearly observable in (b), whereas $O(N)$ of walk ABB is becoming unstable and crosses the zero line in (c). $\overline{O}(N)$, however, remains positive the over entire walk, albeit only slightly above the zero line.
}
\label{fig:3}
\end{figure}

\subsection{Inclusion of phase noise}
\label{BEC-noise}

As mentioned above, the phase angle $\gamma$ was not really controlled in the original experiments reported in \cite{PRL2018, PRA2019, PRR2021}. Hence, it is important to study the impact of phase fluctuations on this parameter. We model the phase noise by a time-dependent variable that is chosen randomly in {\it each step} of the respective walk, i.e.
$\gamma' = \gamma + \delta(N)$ with $\delta(N) \in [-\Delta, \Delta]$, where the $\delta(N)$ are chosen uniformly and independently in the interval. Data in the presence of such a phase noise is shown in Fig. \ref{fig:3}. We see that the winning probabilities of the composite walks ABB remain positive for not too large noise intervals that proves the robustness of our quantum Parrondo game.

\begin{figure}[th]
\centering
\includegraphics[width=0.75\textwidth]{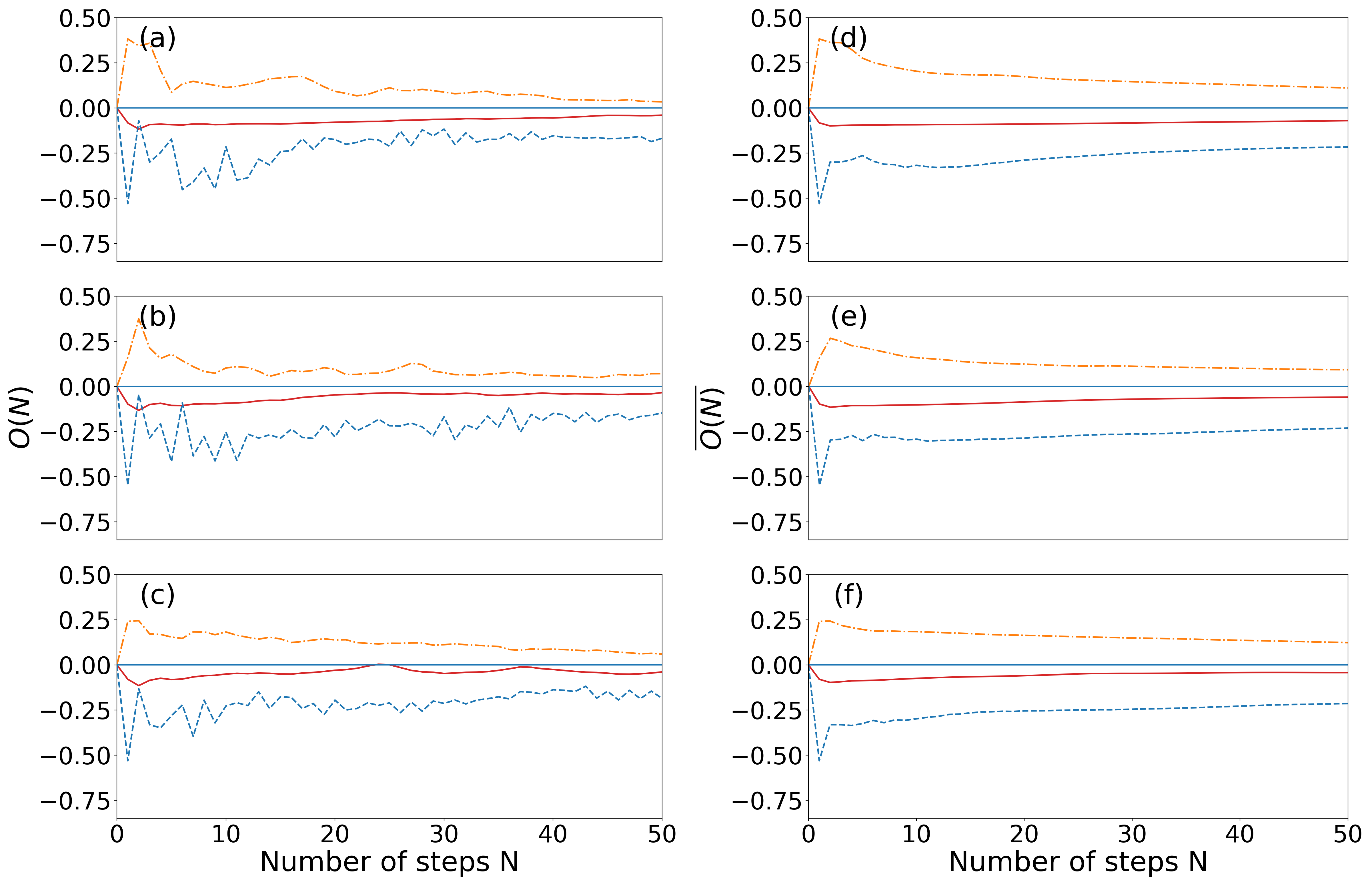}
\caption{Individual winning probabilities $O(N)$ (panels (a) to (c)) and time-averaged $\overline{O(N)}$ (panels (d) to (f)), with the color coding as in Fig. \ref{fig:2}. 
All panels show data with phase noise on $\gamma$ with $\delta(N) \in [- \frac{\pi}{3}, + \frac{\pi}{3}]$. In addition, a quasimomentum distribution is modelled with $\sigma_\beta = 0.005$ (a,d), $\sigma_\beta = 0.01$ (b, e), $\sigma_\beta = 0.02$ in (c, f). All results are averaged over $200$ values of $\beta$.}
\label{fig:4}
\end{figure}

\subsection{Finite quasimomentum distribution}
\label{BEC-QM}

\begin{figure}[th]
\centering
\includegraphics[width=\linewidth]{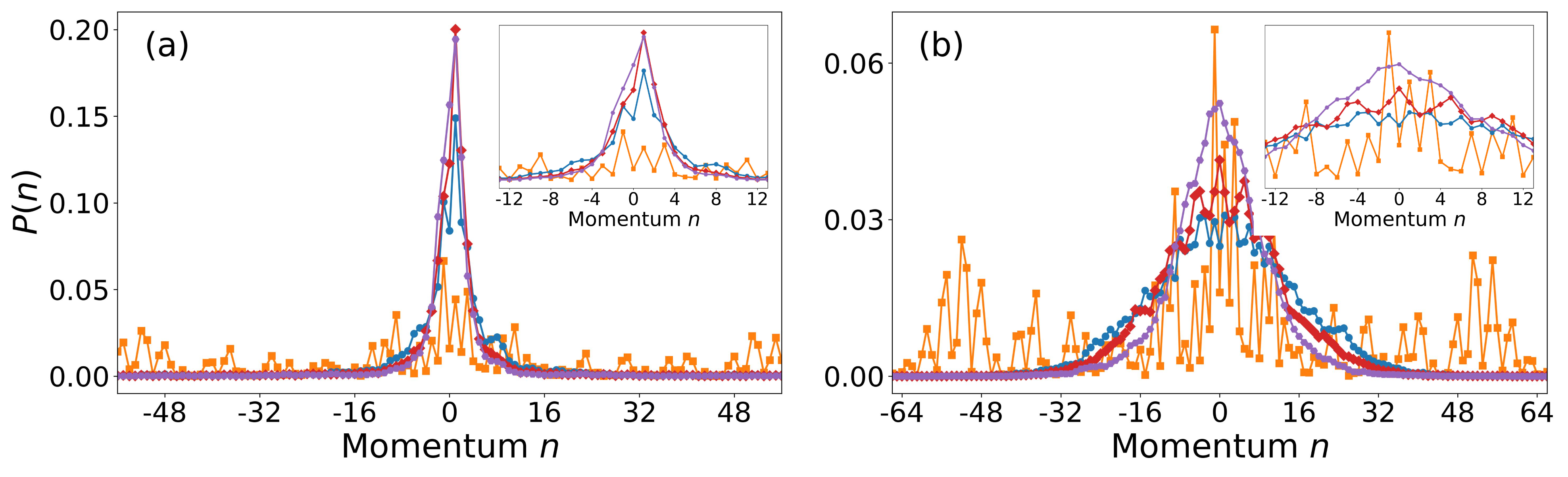}
\caption{(a) Momentum distributions $P(n)$ of the walk ABB after $N = 50$ steps for the resonant kicked-rotor walk ($\beta=0$, orange squares) without noise and with $\sigma_\beta = 0.005, 0.01$ and $0.02$ displayed by the blue circles, red diamonds and purple hexagons, respectively. The right panel (b) shows the corresponding results including a phase noise of $\delta(N) \in [- \frac{\pi}{3}, + \frac{\pi}{3}]$. It appears that only the resonant case has a broader distribution whereas the nonresonant cases are reduced to a peak around $n = 0$ with a certain width and asymmetry (see zoom in the insets). The qualitatively good outcomes of walk ABB in Fig. \ref{fig:4} can be explained by the asymmetry of the central peak, see in particular the inset in (a). The noise broadens the distributions that tends to smear out the overall signal, however, not to such an extent as to lose the asymmetry completely (see Fig. \ref{fig:4}).}
\label{fig:5}
\end{figure}

Up to now, we exclusively considered quantum-resonant dynamics with a fixed quasimomentum $\beta = 0$ and $\tau=4\pi$. Experiments based on Bose-Einstein condensates typically start with an initial momentum distribution with $n=0$ but with a finite width in the Brillouin zone, i.e. with finite quasimomentum. We assume a Gaussian distribution of $\beta$ with zero mean and a standard deviation $\sigma_\beta > 0$. Taking into account a finite quasimomentum induces a coherent phase perturbation due to the free evolution part $\hat F$ of the kicked rotor, see Eq. \eqref{eq:3}. Typically, for small step numbers, the evolution is not sensitive to small nonzero quasimomenta, but deviations occur either for long evolutions or for larger $\beta$ values \cite{WGF2003}. This is confirmed by Fig. \ref{fig:4}. Our results with $\sigma_\beta = 0.005$ remain stable. In both observables, the individual winning probability and the time-average probability, the walk A is no longer a losing game for $\sigma_\beta > 0.01$. However, the composite walk ABB remains a winning game in every single scenario. Comparing to the results from Fig. \ref{fig:2} and Fig. \ref{fig:3}, some of the winning probabilities improved surprisingly when adding coherent phase noise due to the finite quasimomentum distribution. This unexpected behaviour is partly explained by observing directly the momentum distribution of the walk ABB, see Fig. \ref{fig:5}. Here we first see that phase noise and a finite quasimomenta tend to focus the momentum distribution to a peak around $n=0$. Then, this peak broadens with the noise. The central peak, however, remains asymmetric for the walk sequence ABB and hence we obtain a finite positive winning probability. All results were checked as well for larger ensembles with 400 and 800 quasimomenta without any qualitative changes.

\begin{figure}[th]
\centering
\includegraphics[width=0.75\textwidth]{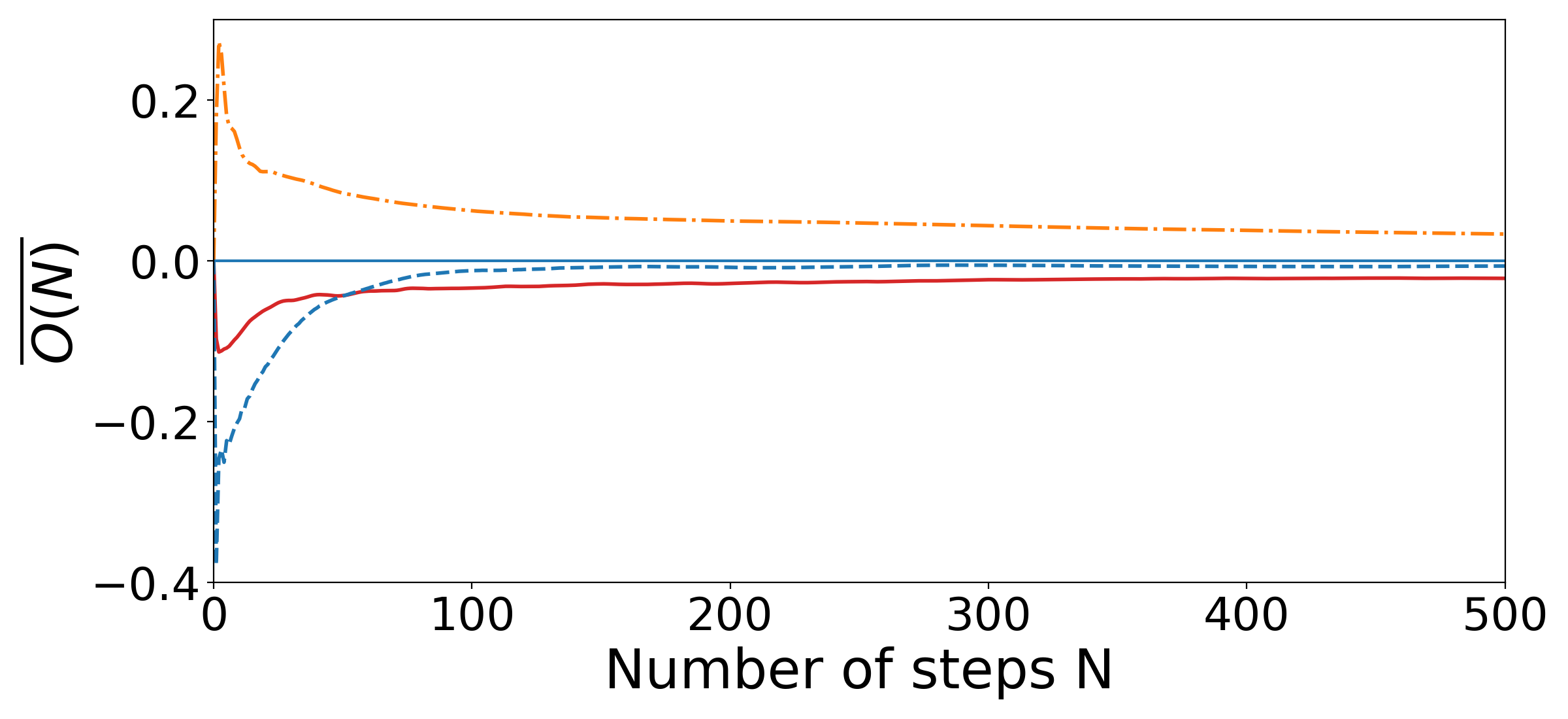}
\caption{The time-averaged winning probability $\overline{O(N)}$ for up to $N = 500$ steps. The coloring of the various walks A, B and ABB is the same as in the previous figures. Noise and quasimomentum distribution as in Fig. \ref{fig:4} (c, f). All the three walks remain stable in $\overline{O(N)}$ and even after 500 steps there is still a nonzero signal. }
\label{fig:6}
\end{figure}

Finally, we tested whether the observed effect of Parrondo's paradox, in principle, survived longer times than just 50 steps. For this, we took the case investigated in Fig. \ref{fig:4} (c, f), and evolved the walks up to 500 steps each, see Fig. \ref{fig:6}. Whilst the single averaged probabilities seem to tend to zero asymptotically, the walks A and B remain losing and simultaneously the sequence ABB remains winning. Hence, the effect indeed survives but seems to fade out for larger and larger step numbers.

\section{Summary}
\label{Sum}

We investigated the adoption of an experiment showing Parrondo's paradox for a sequence of quantum walks. The scenario turns out to be implementable with experiments based on a Bose-Einstein condensates evolving in momentum space. Experimental noise sources, due to phase noise on the MW pulses and arising from the finite quasimomentum width of the condensates, are included. For reasonable perturbations, the effect of the paradox, that while the single walks are losing, the composite walk is winning, survives, also up to a substantial number of walk steps. Hence we are confident that a future experiment as described here may indeed verify Parrondo's paradox in the quantum realm.

\acknowledgments
It is our pleasure to thank F. A. Gr\"unbaum for pointing us to Parrondo's paradox, Nikolai Bolik for numerical support, and the experimental group at Oklahoma State University for helpful discussions.


\begin{thebibliography}{0}

\bibitem{Parrondo1}
J. M. R. Parrondo, How to cheat a bad mathematician, in EEC HC$\&$M Network on
Complexity and Chaos ($\#$ERBCHRX-CT940546 (ISI, Torino, Italy, 1996).

\bibitem{Parrondo2}
J. M. R. Parrondo, G. P. Harmer, D. Abbott, Phys. Rev. Lett., {\bf 85}, 5226 (2000)

\bibitem{Harmer1}
G. P. Harmer, D. Abbott, Nature {\bf 402}, 864 (1999)

\bibitem{Harmer2}
G. P. Harmer, D. Abbott, Stat. Sci. {\bf 14}, 206 (1999)

\bibitem{CQP-review}
J. W. Lai and K. H. Cheong,
Nonlinear Dynamics {\bf 100}, 849 (2020); and refs. therein

\bibitem{CP0}
A. Allison, D. Abbott,  Chaos: An Interdisciplinary Journal of Nonlinear Science {\bf 11}, 3 (2001)

\bibitem{CP1}
J. M. R. Parrondo, L. Dinís, Contemporary Physics, {\bf 45}, 2 (2004)

\bibitem{Portugal}
R. Portugal, Quantum Walks and Search Algorithms (Springer, New York, 2013)

\bibitem{QRW}
Y. Aharonov, L. Davidovich, N. Zagury, Phys. Rev. A, {\bf 48}, 1687 (1993)

\bibitem{QP1}
C.M. Chandrashekar and S. Banerjee, Phys. Lett. A {\bf 375} 1553 (2011) 

\bibitem{QP2}
J. Rajendran and C. Benjamin, EPL {\bf 122},  40004 (2018);
R. Soc. Open Sci. {\bf 5}, 171599 (2018)

\bibitem{QP3}
A. P. Flitney, arXiv preprint arXiv:1209.2252 (2012)

\bibitem{Ratchet}
L. Chen, C.-F. Li, M. Gong, and G.-C. Guo, Physica A {\bf 389}, 4071 (2010)

\bibitem{Game}
J. Ng and D. Abbott, Introduction to quantum games and a quantum Parrondo game, in Advances in Dynamic Games {\bf }, 
Nowak A.S., Szajowski K. (eds),  Annals of the International Society of Dynamic Games, vol 7. (Birkh\"auser Boston, 2005), p. 649.


\bibitem{AdvQT}
M. Jan, Q. Q. Wang, X. Y. Xu, W. W. Pan, Z. Chen, Y. J. Han, C. F. Li, G. C. Guo,
D. Abbott, Adv. Quantum Technol. {\bf 3}, 1900127 (2020).

\bibitem{QP0}
A. P. Flitney, D. Abbott, and N. F. Johnson, J. Phys. A: Math. Gen. {\bf 37} 7581 (2004)

\bibitem{FNL}
M. Li, Y. S. Zhang, G. C. Guo,
Fluctuation and Noise Letters {\bf 12}, 1350024 (2013)

\bibitem{LMP2016}
F. A. Gr\"unbaum, M. Pejic, Lett. Math. Phys. {\bf 106}, 251 (2016)

\bibitem{PRL2018}
S. Dadras, A. Gresch, C. Groiseau, S. Wimberger, G. S. Summy,
Phys. Rev. Lett. {\bf 121}, 070402 (2018)

\bibitem{PRA2019}
S. Dadras, A. Gresch, C. Groiseau, S. Wimberger, and G. S. Summy, 
Phys. Rev. A {\bf 99}, 043617 (2019)

\bibitem{PRR2021}
J. H. Clark, C. Groiseau, Z. N. Shaw, S. Dadras, C. Binegar, S. Wimberger, G. S. Summy, and Y. Liu, 
Phys. Rev. Res. {\bf 3}, 043062 (2021)

\bibitem{LC2021}
J. W. Lai and K. H. Cheong,
Phys. Rev. Res. {\bf 3},  L022019 (2021); and refs. therein

\bibitem{WB2021}
Z. Walczak and J. H. Bauer,
Phys. Rev. E {\bf 104}, 064209 (2021)

\bibitem{AdvAtMol}
M. Sadgrove, S. Wimberger, Adv. At. Mol. Opt. Phys. {\bf 60}, 315 (2011)

\bibitem{PhysRep90}
F. M. Izrailev, Phys. Rep. {\bf 196},  299 (1990)

\bibitem{PhysRep22}
M. Santhanam, S. Paul, J. B. Kannan,  Phys. Rep. {\bf 956}, 1 (2022)

\bibitem{PRA2016a}
G. Summy and S. Wimberger, Phys. Rev. A {\bf 93}, 023638 (2016)

\bibitem{PRA2016}
J. Ni, W. K. Lam, S. Dadras, M. F. Borunda, S. Wimberger, G. S. Summy,
Phys. Rev. A {\bf 94}, 043620 (2016)

\bibitem{rsos}
J. Rajendran, C. Benjamin, R.Soc.opensci. {\bf5}, 171599 (2018)

\bibitem{AP2017}
J. Ni, S. Dadras, W. K. Lam, R. K. Shrestha, M. Sadgrove, S. Wimberger, and G. S. Summy 
 Ann. Phys. {\bf 529} (8), 1600335 (2017)

\bibitem{Mark2007}
M. Sadgrove, M. Horikoshi, T. Sekimura, and K. Nakagawa, Phys. Rev. Lett. {\bf 99}, 043002 (2007);
I. Dana, V. Ramareddy, I. Talukdar, and G. S. Summy, Phys. Rev. Lett. {\bf 100}, 024103 (2008)

\bibitem{Walk-Selection}
M. Karski, L. F\"orster, J.-M. Choi, A. Steffen, W. Alt, D. Meschede, and A. Widera, Science {\bf 325}, 174 (2009);
 H. B. Perets, Y. Lahini, F. Pozzi, M. Sorel, R. Morandotti, and Y. Silberberg, Phys. Rev. Lett. {\bf 100}, 170506 (2008);
H. Schmitz, R. Matjeschk, C. Schneider, J. Glueckert, M. Enderlein, T. Huber, and T. Schaetz, Phys. Rev. Lett. {\bf 103},
090504 (2009);
A. Schreiber, K. N. Cassemiro, V. Potocek, A. Gabris, P. J. Mosley, E. Andersson, I. Jex, and C. Silberhorn, Phys. Rev.
Lett. {\bf 104}, 050502 (2010);
L. Sansoni, F. Sciarrino, G. Vallone, P. Mataloni, A. Crespi, R. Ramponi, and R. Osellame, Phys. Rev. Lett. {\bf 108}, 010502 (2012);
F. Cardano, A. D'Errico, A. Dauphin, M. Maffei, B. Piccirillo, C. de Lisio, G. De Filippis, V. Cataudella, E. Santamato, L. Marrucci, M. Lewenstein, and 
P. Massignan, Nat. Commun. {\bf  8}, 15516 (2017)


\bibitem{ChinPhysB}
M. Li, Y. S. Zhang, G. C. Guo,
Chin. Phys. B {\bf 22}, 030310 (2013)

\bibitem{PRA2015}
M. Weiss, C. Groiseau, W. Lam, R. Burioni, A. Vezzani, G. Summy, and S. Wimberger,
Phys. Rev. A {\bf 92}, 033606 (2015)

\bibitem{PRA2019b}
C. Groiseau, A. Gresch, and S. Wimberger, J. Phys. A: Mathematical and Theoretical {\bf 51}, 275301 (2018)

\bibitem{WGF2003}
S. Wimberger, I. Guarneri, and S. Fishman, Nonlinearity {\bf 16}, 1381 (2003) 

\end{thebibliography}
\end{document}